# Two-dimensional metal in a parallel magnetic field

Xuan P. A. Gao[1,2], Allen P. Mills, Jr.[1], Arthur P. Ramirez[3], Loren N. Pfeiffer[1], Kenneth W. West[1]

[1]*Bell Laboratories, Lucent Technologies, Murray Hill, New Jersey 07974*
[2]*Dept. of Applied Physics & Applied Math, Columbia University, New York City, NY 10027*
[3]*Los Alamos National Laboratory, Los Alamos, NM 87545*

We have investigated the effect of an in plane parallel magnetic field ($B_\parallel$) on two high mobility metallic-like dilute two-dimensional hole gas (2DHG) systems in GaAs quantum wells. The experiments reveal that, while suppressing the magnitude of the low temperature resistance drop, $B_\parallel$ does not affect $E_a$, the characteristic energy scale of the metallic resistance drop. The field $B_c$ at which the metallic-like resistance drop vanishes is dependent on both the width of quantum well and the orientation of $B_\parallel$. It is unexpected that $E_a$ is unaffected by $B_\parallel$ up to $B_c$ depite that the Zeeman energy at $B_c$ is roughly equal to $E_a$.

According to the scaling theory of localization, a non-interacting two-dimensional (2D) fermion system is localized at zero temperature and magnetic field [1]. Therefore, after the discovery of a metal-insulator transition (MIT) in a 2D electron gas in a clean Si-MOSFET [2], the nature of the 2D metallic state has been a subject of much debate [3]. Experimentally it has been found that the metallic resistance per square $R_{xx}$ is thermally activated with an activation energy $E_a$ around 10% of the Fermi energy, $E_F$ [4-6]. We note that Coulomb interactions are most pronounced in high mobility systems with high values of the Wigner-Seitz radius, $r_s$, the dimensionless measure of the strength of the inter-particle Coulomb interaction in units of the Fermi energy. This has led to the theoretical proposal that the 2D system is driven into the metallic state by strong Coulomb interactions [7]. On the other hand, it has been argued that the metallic like behavior is caused by classical temperature dependent disorder scattering or temperature dependent screening [8]. It has also been found that an in plane magnetic field induces a giant positive magneto-resistance and drives the system from a metallic state ($dR_{xx}/dT>0$) to an insulating state ($dR_{xx}/dT<0$) for $B_\parallel$ greater than a 'critical field' $B_c$ for which $dR_{xx}/dT=0$, suggesting that the spins of the carriers may play an important role since $B_\parallel$ mainly couples to the spins of the carriers [9-11]. Many explanations have been suggested, ranging from gas-liquid phase separation [12] and different kinds of superconductivity to single particle physics descriptions based on magnetic field driven disorder [13], coupling of $B_\parallel$ to the orbital motion [14], or classical percolation [15].

While the parallel magnetic field effect has been systematically investigated mostly in n-Si-MOSFET's [3], measurements are especially interesting on GaAs/AlGaAs quantum well samples, where the mobility of the carriers is typically much higher. Our experiments demonstrate that for $B_\parallel < B_c$, $B_\parallel$ only suppresses the size of the metallic resistance drop while the energy scale $E_a$ is invariant. When $B_\parallel > B_c$, the resistance of the 2D system keeps increasing slowly upon lowering the temperature, in contrast to the strongly insulating behavior in n-Si-MOSFET's. We find that $B_c$ is anisotropic with respect to the orientation of $B_\parallel$ in the 2DHG plane and depends on the width of the quantum well (QW). This can be partly explained by the known anisotropic g factor in GaAs QW's. The field independence of the characteristic energy scale, $E_a$ has never been reported and constraints theoretical models describing the origin of the metallic-like resistance drop.

Our transport measurements were performed down to as low as 10mK on two high mobility low density 2DHG in GaAs quantum wells with well widths of 10 and 30nm. Since we have noticed the heating effect of measurement powers higher than a few fW/cm$^2$ at low temperature in a previous study [16], we have consistently used measurement signal powers less than 3fW/cm$^2$. The 30nm wide QW sample has density of $1.03\times10^{10}$ cm$^{-2}$, and low temperature hole moblility, $\mu_p = 7.5 \times 10^5$ cm$^2$V$^{-1}$s$^{-1}$. This very same sample was previously used in obtaining one of the data sets in [6]. The 10nm well sample has nearly the same density, $1.14\times10^{10}$ cm$^{-2}$, and low temperature hole moblility of $3.4 \times 10^5$ cm$^2$ V$^{-1}$ s$^{-1}$. The samples were grown on (311)A GaAs wafers using Al$_x$Ga$_{1-x}$As barriers (typical x = 0.10) and symmetrically placed delta-doping layers above and below the pure GaAs QW's. The samples were prepared in the form of Hall bars, of approximate dimensions (2.5×9) mm$^2$, with diffused In(5% Zn) contacts. The measurement current (~ 100pA, 4Hz) was applied along the [$\bar{2}$33] direction in all our experiments. Independent measurements of the longitudinal resistance per square, $R_{xx}$, from contacts on both sides of the sample were made simultaneously as the temperature or applied magnetic field was varied. The samples were mounted in a top-loading dilution refrigerator. The temperature was read from a Ge resistance thermometer attached to the refrigerator mixing chamber. The Ge thermometer was calibrated down to 4mK by He-3 melting curve thermometry. Hall resistance measurements were used to determine θ, the angle at which $B_\parallel$ was tilted from the 2DHG plane.

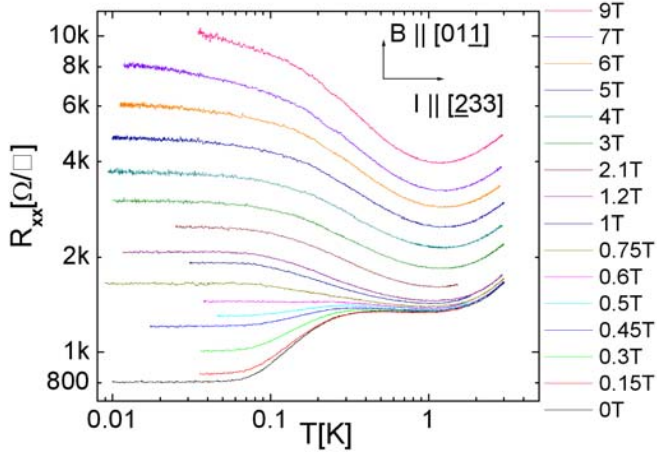

FIG. 1. $R_{xx}$ vs. T at different parallel magnetic field $B_\parallel$ of 2DHG in a 30nm wide p-GaAs/AlGaAs quantum well. The density of the 2DHG is $1.03 \times 10^{10}$ cm$^{-2}$. $B_\parallel$ is tilted from the 2DHG plane with angle $\theta \approx 0.8^\circ$.

In Fig. 1, the T dependence of $R_{xx}$, the longitudinal resistance per square, of the 30nm QW at various $B_\parallel$'s is shown on a log-log plot. The in plane parallel magnetic field $B_\parallel$ was perpendicular to the current, i.e. in the [01$\bar{1}$] direction. The high temperature data are all dominated by Bloch-Gruneisen phonon scattering, which has an asymptotic $T^3$ dependence at low temperatures and a linear dependence on T at high temperatures. This phonon contribution can be approximated by $R_{ph}(T) = R_{ph}t^3/(1+t^2)$, where $t=T/T_{ph}$ [17]. It is obvious from Fig. 1 that the phonon term also has a $B_\parallel$ dependence - $R_{ph}$ increases monotonically with $B_\parallel$ and $T_{ph}$ stays constant. We shall focus our discussion on the data for T<1K where the metallic like resistance drop emerges and the phonon scattering is negligible. There are two most striking features in Fig. 1. First, when $B_\parallel < B_c \sim 0.55T$, all the $R_{xx}(T)$ traces exhibit $dR_{xx}/dT \geq 0$ below ~0.3K. Furthermore, clearly the in plane magnetic field $B_\parallel$ suppresses the magnitude of the resistance drop, while $B_\parallel$ has a much smaller effect above 0.3K. Second, when $4B_c > B_\parallel > B_c$, $R_{xx}$ never exhibits $dR_{xx}/dT>0$ over all the temperature range T<1K but instead saturates below a temperature $T_s$. The saturation temperature $T_s$ systematically decreases as $B_\parallel$ increases. We note that this saturation of resistance at low T and low field is unlikely to be caused by heating caused by the measuring signal or environmental RF noise. We used ~3fW/cm$^2$ driving power and self-contained well shielded electronics in all the measurements, under which conditions we have shown the ability to cool our 2DHG to less than 10mK [16]. This is supported by the continued temperature dependence of $R_{xx}$ below 20mK at high $B_\parallel$. The saturation of $R_{xx}$ at low T above $B_c$ is in contrast to the strongly insulating behavior in Si-MOSFET's, which is usually characterized by variable range hopping conduction such that $R \sim R_0 \exp(\Delta/T)^n$. We note here that our sample resistivity is much lower than $h/e^2$, the typical resistivity of Si-MOSFET samples where the parallel field causes a dramatic effect and drives the 2D system insulating. Therefore, the strongly insulating behavior observed in the Si-MOSFET's in a parallel field may be partly caused by disorder, such as suggested in [13]. Moreover, as $B_\parallel$ increases well above $4B_c$, the sample is either gradually entering a weakly insulating state in which $R_{xx}$ tends to rise slowly at low T, or $T_s$ is below our measurement range. In this paper we will concentrate on the region $B_\parallel < B_c$. The physics at $B_\parallel > B_c$ needs further investigation.

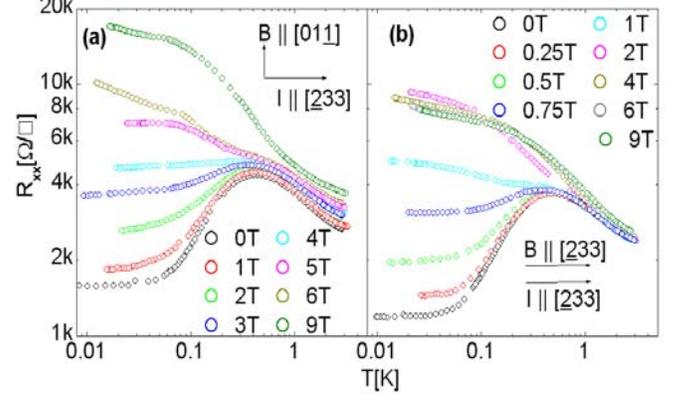

FIG. 2. $R_{xx}$ vs. T at various $B_\parallel$ for the 10nm wide quantum well. The density of the 2DHG is $1.14 \times 10^{10}$ cm$^{-2}$. (a) $B_\parallel$ along [01$\bar{1}$] direction, $B_\parallel$ is tilted from 2DHG plane with angle $\theta \approx 2^\circ$. (b) $B_\parallel$ along [$\bar{2}$33] direction, angle $\theta \approx 0.1^\circ$. Sample resistance at zero field was changed between (a) and (b) due to thermal recycle of sample from 10mK to room temperature.

Our measurements of $R_{xx}(T)$ at different $B_\parallel$ for the 10nm wide QW sample are presented in Fig. 2. The behavior for $B_\parallel<B_c$ is qualitatively similar to that in Fig.1. We note that $B_\parallel$ was tilted from the 2DHG plane only 0.1$^\circ$ for Fig.2b, the most parallel case in our experiments. For $B_\parallel >B_c$ there is a nonmonotonic magneto-resistance up to 9T at low temperature. The different behaviors at $B_\parallel>B_c$ on Fig. 1 and Fig. 2 hint that the low temperature transport property of the system at $B_\parallel>B_c$ may be sensitive to small perpendicular field and it is currently being studied. For Fig.2a $B_\parallel$ was applied along the [01$\bar{1}$] direction, $B_c$ is roughly 4 T. When $B_\parallel$ is changed to the [$\bar{2}$33] direction, $B_c$ changes to roughly 0.9T. The change of $B_c$ correlates quantitatively with the known anisotropy of the g factor in GaAs QW's [18]. Calculations in [18] show that the g factor when $B_\parallel$ is along [$\bar{2}$33], $g_{[\bar{2}33]} \approx 4g_{[01\bar{1}]}$, where $g_{[01\bar{1}]}$ is the g factor when $B_\parallel$ is along [01$\bar{1}$]. This implies that the suppression of metallic resistance is driven by the Zeeman spin splitting energy. It is, however, surprising that $B_c$ differs by about factor of 7 for the 10nm and 30nm well widths, although $E_a$ is substantially the same. A natural interpretation would be that the g factor is bigger in the wider QW.

To quantitatively characterize the in plane magnetic field effect on the metallic resistance drop, we fitted the T <0.2K and $B_\parallel < B_c$ part of all the data by

$$R_{xx}(T) = R_0 - R_a[1-\exp(-E_a/k_BT)] \quad (1)$$

The second term on the right side of Eq.(1) accounts for the anomalous metallic-like resistance drop with magnitude $R_a$. We found that $E_a = (0.246 \pm 0.021)$ K for Fig.1 and Fig. 2b and the zero field data of Fig. 2a, i.e. $E_a$ is not affected by $B_\parallel$ less than $B_c$. The shapes of the resistance drop curves in Fig. 2a at nonzero field differ from the other curves and would imply $E_a<0.246K$. We believe these smaller values of $E_a$ are an artifact associated with the effect of the relatively large perpendicular component of $B_\parallel$. We estimated $B_\parallel$ was tilted with the 2DHG plane from 0.8°, 2° and 0.1° for Fig.1, Fig.2a and Fig.2b respectively, using the Hall resistance as a measure of the perpendicular field. Here the parallel field independence of $E_a$ is in contrast to the linearly decreasing $E_a$ with $B_\parallel$ in an n-Si-MOSFET as reported in [10]. In Fig. 3a we show the normalized metallic resistance drop, $(R_{xx}(B_\parallel, T) - R_0(B_\parallel))/R_a(B_\parallel)$ for the three sets of data, where the parameters $R_0$ and $R_a$ were extracted from fitting the data by Eq. (2) with $E_a=0.246K$. The normalized $T<E_a=0.246K$ data on Fig.3a collapse on the function $\exp(-0.246\text{Kelvin}/T)-1$, which thus characterizes the shape of the metallic resistance drop for $B_\parallel<B_c$. The normalized magnitude of the resistance drop, $R_a(B_\parallel)/R_a(0)$ is plotted versus the normalized parallel field $B_\parallel/B_c$ in Fig.3b, which suggests that there is a universal quadratic parallel field dependence of the magnitude of the metallic resistance drop for our samples. The normalized $R_0$, $R_0(B_\parallel)/R_0(0)$ has a much weaker dependence on $B_\parallel/B_c$.

Evaluation of the Zeeman energy at $B_c$, $\Delta E=g\mu_B B_c$, suggests it is related to the activation energy $E_a$ rather than the Fermi energy $E_F$. If we assume $\Delta E=E_a=0.246K$, we find $g_{[01\underline{1}]}=0.67$ for the 30nm QW and $g_{[01\underline{1}]}=0.09$, $g_{[2\underline{3}3]}=0.40$ for the 10nm QW. The possibility that $\Delta E$ is the Zeeman energy required to fully polarize the 2D hole gas ($\Delta E = 2E_F$ for noninteracting particles) is not reasonable because the implied values of g would be of an order of magnitude bigger, since $E_F=1.5K\sim6E_a$.

The magneto-resistance $R_{xx}(B_\parallel)$ traces for the 30nm QW at various temperatures are shown in Fig. 4. The detailed evolution of $R_{xx}(B_\parallel)$ vs. T shows that at temperatures lower than 0.2K a steep $R_{xx}$ vs. $B_\parallel$ emerges in the $B_\parallel < 1$Tesla region. This corresponds to the suppression of the metallic-like resistance drop by the same parallel field as in Fig. 1. For $T \geq 1K$, in the $B_\parallel <$ 5Tesla region $R_{xx}(B_\parallel)$ follows the form $\exp(B_\parallel^2)$, while $R_{xx}(B_\parallel) \propto \exp(B_\parallel)$ as $B_\parallel > 5$Tesla. This different type of magneto-resistance at high field and low field might be caused by the coupling of $B_\parallel$ with orbital motion as a result of the finite well width, as suggested in [14].

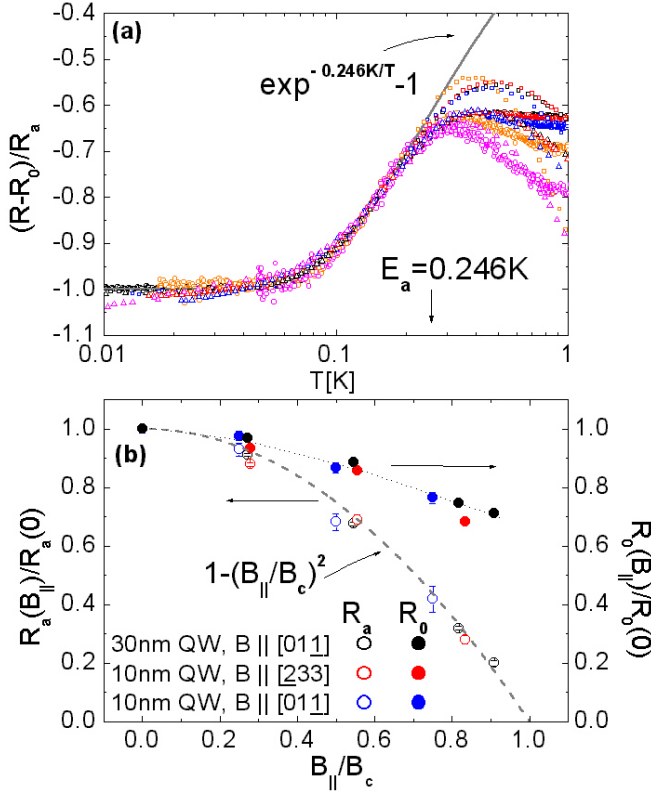

FIG. 3. (a) Normalized metallic resistance drop for $B_\parallel<B_c$ by procedure described in text. Open circles are from the data on Fig. 1, open triangles are from data on Fig. 2a and open squares are from data on Fig. 2b. The solid gray line shows the function $\exp(-0.246\text{Kelvin}/T)-1$. (b) Normalized magnitude of the metallic resistance drop, $R_a(B_\parallel)/R_a(B_\parallel=0)$ vs. $B_\parallel/B_c$ and $R_0(B_\parallel)/R_0(B_\parallel=0)$ vs. $B_\parallel/B_c$; the corresponding values of $B_c$ are 0.55±0.01, 4.02±0.25 and 0.89±0.05 T, where the error estimates indicate the range of $B_\parallel$ for which the $T\leq0.3K$ isothermal magnetoresistance curves cross. The dashed gray line is the function $y=1-x^2$, with no adjustable parameters. The dotted black line connecting scaled $R_0$ is a guide to eye.

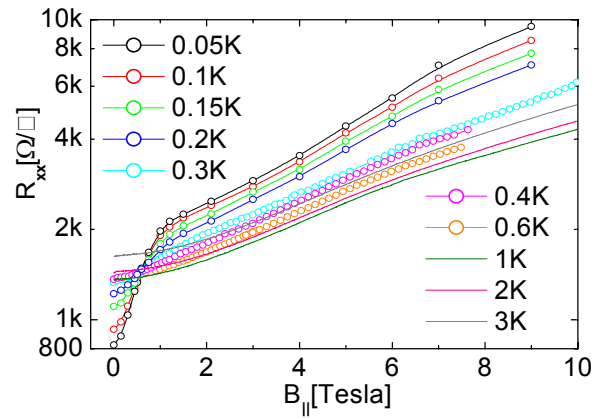

FIG. 4. $R_{xx}$ vs. $B_\parallel$ along $[01\underline{1}]$ at different temperatures of the 30nm wide QW.

Phillips *et al.* argued that the drop of resistivity at low T is caused by the superconductive pairing of carriers [19]. Based on that model, $E_a$ would be related to the mean field critical temperature at which Cooper pairs start forming and the saturation of resistance at low temperature is attributed to the lack of global phase coherence of Cooper pairs caused by disorder effects and the low superfluid density in [19]. Our finding that $E_a$ is independent of $B_\parallel$ does not appear to agree with the simplest superconductivity interpretation in which $E_a$ would be expected to decrease as $B_\parallel$ increases.

An interesting observation is that the existence of two spin subbands can cause an Arrhenius temperature dependence of resistance [20]. It is also observed that the magnitude of the metallic-like resistance increases with the magnitude of the spin splitting [21]. In our samples, spin-orbit interactions are minimized by doping symmetrically from both side of the QW. We estimate the residual spin splitting from spin orbital coupling at zero field is negligible. In the presence of $B_\parallel$, the holes are split into two spin bands. The decreasing magnitude of the metallic resistance drop with $B_\parallel$ of our samples is not in agreement with a spin splitting origin picture of the 2D metallic behavior.

A satisfactory model explaining the 2D metallic phenomena in p-GaAs should predict both a field independent $E_a$ and the apparent quadratic dependence of the low temperature metallic resistance drop of the Zeeman energy. Among various available theoretical models, a gas-liquid condensation picture driven by strong Coulomb correlations is a possibility [12]. In such a theory the field insensitivity of $E_a$ would be associated with the condensation temperature depending only on inter-particle Coulomb energy. The effect of a parallel field would be to vaporize some of the liquid phase, thus leading to the observed resistance increasing.

In summary, we have studied the parallel magnetic field effects on two high mobility dilute 2DHG in GaAs quantum wells. $B_c$, the field which fully suppresses the metallic resistance drop of the system at low temperatures is found to be dependent on the orientation of $B_\parallel$ and the width of quantum well. Meanwhile, the observations of a field independent $E_a$ and a universal quadratic parallel magnetic field dependence of the size of metallic resistance drop set strong constraints on theoretical models of the 2D metallic state.

We gratefully acknowledge discussions with S. H. Simon, C. M. Varma and D. C. Tsui.

---